\documentclass{appolb}
\usepackage{epsfig}

\begin{document}

\title{Percolation cluster formation at ultrarelativistic heavy ion
collisions}

\date{\today}

\author{M.K. Suleymanov\thanks{%
E-mail: mais$\_$suleymanov@comsats.edu.pk}, E. U. Khan, K. Ahmed, Mahnaz Q.
Haseeb, Farida Tahir and Y. H. Huseynaliyev 
\address{Department of Physics \  COMSATS Institute of Information Technology,
Islamabad, Pakistan}}
\title{}
\maketitle

\begin{abstract}
We expect that the experimental study of percolation cluster formation and
appearance of the critical transparency of the strongly interacting matter
can give the information about the onset state of deconfinement.
\end{abstract}

%\title{The Paper Title comes Here...\thanks{}}

\eqsec  % uncomment this line to get equations numbered by (sec.num)

\pagestyle{plain} 
%% uncomment the following line to get equations numbered by (sec.num)
\eqsec
\newcount\eLiNe\eLiNe=\inputlineno\advance\eLiNe by -1

\section{Introduction}

The deconfinement in ultrarelativistic collisions is expected when the
density of quarks and gluons becomes so high that due to strong overlap, it
no longer makes sense to partition them into color-neutral hadrons. The
clusters get much larger than hadrons, within which color is not confined;
deconfinement may be, therefore, related to cluster formation~\cite{[1]}. In
Ref.~\cite{[2]} we have discussed that the regime change which is indicated
in the behavior of some characteristics of the particles production at
ultrarelativistic heavy ion collisions as a function of centrality (e.g.,~%
\cite{[3]}) could be explained with the cluster formation as a result of the
nucleon and quark percolation~\cite{[4]}. Experimental observation of the
effects connected with formation and decay of the percolation clusters in
heavy ion collisions at ultrarelativistic energies and the study of
correlation between these effects could provide the information about
deconfinement of strongly interacting matter in clusters. To confirm the
deconfinement in cluster it is necessary to study the centrality dependence
in the behavior of secondary particles yields and simultaneously, critical
increase in transparency of the strongly interacting matter.

Appearance of the critical transparency could change the absorption
capability of the medium and we may observe a change in the heavy flavor
suppression depending on their kinematical characteristics. So, study the
centrality dependence of heavy flavor particle production with fixed
kinematical characteristics could give the information on changing of
absorption properties of medium depending on the kinematical characteristics
of heavy flavor particles.

\section{Experimental Analysis}

A comparison of yields in different ion systems by using nuclear
modification factors such as R$_{CP}$ (involving Central and Peripheral
collisions e.g . for \ heavy flavor particles yields) could provide
information on the properties of the nuclear matter. In such definition of ,
appearance of transparency could be identified and detected using the
condition .Using some statistical and percolation models~\cite{[8]} and
experimental data on the behaviour of the nuclear modification factors it is
possible to get information on the appearance of the anomalous nuclear
transparency as a signal of formation of the percolation cluster. Recent
data obtained by STAR RHIC BNL ~\cite{[6]} on the behavior of the nuclear
modification factors of the strange particles as a function of the
centrality in Au+Au- and p+p-collisions at $\sqrt{s_{NN}}=200GeV$ may help
us to answer the questions: how the new phases of strongly interacting
matter form? May we expect a signal on the formation of the intermediate
nuclear system e.g., nuclear cluster? The strange particles could be formed
as a result of quark coalescence in high density strongly interacting matter
and on other hand they could be captured by this system intensively. So by
increasing the centrality, yields of heavy flavors could decrease. The
appearance of superconducting property of the strongly interacting matter~%
\cite{[9]} as a result of the formation of percolation cluster could stop
decrease of yields of heavy flavors. 

%%%%%%\bigskip \FRAME{ftbpF}{1.8775in}{2.2286in}{0pt}{}{}{Figure}{\special%
%%%%%%{language "Scientific Word";type "GRAPHIC";maintain-aspect-ratio
%%%%%%TRUE;display "USEDEF";valid_file "T";width 1.8775in;height 2.2286in;depth
%%%%%%0pt;original-width 1.446in;original-height 1.721in;cropleft "0";croptop
%%%%%%"1";cropright "1";cropbottom "0";tempfilename
%%%%%%'JJXDOZ00.wmf';tempfile-properties "XPR";}}

\end{document}